# The low-energy positron scattering upon endohedrals

## M. Ya. Amusia[1;21] and L. V. Chernysheva[2]

[1]*The Racah Institute of Physics, the Hebrew University, Jerusalem 91904, Israel*
[2]*A. F. Ioffe Physical-Technical Institute, St. Petersburg 194021, Russian Federation*

**Abstract:**

We investigate positron scattering upon endohedrals and compare it with electron-endohedral scattering. We show that the polarization of the fullerene shell considerably alters the polarization potential of an atom, stuffed inside a fullerene. This essentially affects both the positron and electron elastic scattering phases as well as corresponding cross-sections. Of great importance is also the interaction between the incoming positron and the target electrons that leads to formation the virtual positronium $\tilde{P}s$. We illustrate the general trend by concrete examples of positron and electron scattering upon endohedrals $He@C_{60}$ and $Ar@C_{60}$, and compare it to scattering upon fullerene $C_{60}$.

To obtain the presented results, we have employed new simplified approaches that permit to incorporate the effect of fullerenes polarizability into the $He@C_{60}$ and $Ar@C_{60}$ polarization potential and to take into account the virtual positronium formation. Using these approaches, we obtained numeric results that show strong variations in shape and magnitudes of scattering phases and cross-sections due to effect of endohedral polarization and $\tilde{P}s$ formation.

PACS numbers: 34.80.Bm, 34.80.Gs, 36.10.Dr

**1**. At first glance, the positron $e^+$ scattering upon an endohedral one can treat similarly to electron $e^-$ scattering [1-3], simply by neglecting the exchange between the projectile and the target electrons. No doubt that the endohedral polarization that proved to be important in the low-energy $e^- + A@C_{60}$ [1-3] plays also an important role in $e^+ + A@C_{60}$ collisions.

However, it was understood already long ago that in treating the $e^+ + A$ scattering, one has to take into account the relatively strong interaction between $e^+$ and the atomic electrons. This interaction impressively modifies the motion of the $e^+e^-$ pair in the field of the residual ion $A^+$. Since the Hartree potential of $A$ that acts upon $e^+$ is repulsive, $e^+$ concentrates in the same areas as the virtually exited atomic electrons. They can form temporarily a virtual positronium $\tilde{P}s$ [4, 5] that dramatically affects the $e^+$ scattering cross-section. Our aim here is to check whether similar effect is of importance in low-energy elastic $e^+ + A@C_{60}$ scattering.

Thus, to calculate $e^+ + A@C_{60}$ scattering, two major effects has to be taken into account simultaneously, namely the polarizability of the target endohedral and the effect of $\tilde{P}s$ formation. We also have to estimate the alteration of the endohedral potential, when one goes from $e^-$ to $e^+$ as a projectile.

---

[1] amusia@vms.huji.ac.il

We discussed already in [1, 2] the amazing result that an addition of a single relatively small atom inside a fullerene affects essentially the electron elastic scattering cross-section of the latter, in spite the fact that the presence of an additional atom inside alters negligibly the total size of the target system under consideration. As it has been demonstrated in [1-3] that the quantum interference changes the situation impressively, so that the total phase $\delta_l^{A@C_N,-}$ of the partial wave $l$ of an electron scattered upon endohedral A@$C_N$ is with good accuracy equal to the sum of scattering phases $\delta_l^{A,-}$ and $\delta_l^{C_N,-}$ of an electron upon atom A, stuffed inside the fullerene $C_N$, and the $C_N$ itself. This was dubbed as a phase additivity rule. It means, counterintuitively, that for $e^-$ scattering a single atom contribution is quite big as compared to the background of $C_N$ cross-section.

We have performed calculations for $e^-$ scattering, assuming that it feels the Hartree-Fock $\hat{V}_{HF}(r)$ potential of the atom A, as well as the $e^-$ static $W_F^-(r)$ and polarization $V_F^{pol,-}(r)$ potentials of the $C_N$. The inclusion of $V_F^{pol,-}(r)$ proved to be very important, since $C_N$ for N>>1 is a highly polarizable object, as compared to the atom $A$.

We took into account the effect of the polarization of the atom and the mutual $C_N$ and $A$ polarizations that was called polarization interference. We have demonstrated in [2, 3] that this is a big effect in $e^-$ scattering.

The aim of this Letter is to investigate $e^+$ scattering upon endohedrals and to see, what role in this processes plays the atomic Hartree potential, the atomic polarization potential $V_A^{pol,+}$, the static potential of the fullerene $W_F^+(r)$ and its polarization potential $V_F^{pol,+}(r)$ as well as the polarization interference $V_{FA}^{pol,+}(r)$. We will compare all the results for $e^+$ with that for $e^-$.

As concrete objects of calculations we choose almost ideally spherical fullerene $C_{60}$ and endohedrals $He@C_{60}$, $Ar@C_{60}$ along with the empty fullerene $C_{60}$. As $A$ we consider quite small and spherical atoms $He$ and $Ar$ that are centrally located in considered endohedrals.

In studies of elastic scattering of both $e^+$ and $e^-$ it is essential to have in mind some important difference in general behavior of their respective phase shifts. Let the phases $\delta_l(E)$ as functions of energy $E$ be normalized in such a way that $\delta_l(E \to \infty) \to 0$. Since the considered targets consist of electrons and nuclei, for $e^+$ there is no exchange between the projectile and the target constituents. Therefore the following expression is valid $\delta_l^+(0) = n_l^+ \pi$, where $n_l^+$ is the total number of bound states in the system "positron +endohedral" [6]. The presence of exchange in $e^- + A@C_N$ collisions leads to an essentially different relation for the $e^-$ scattering phases at zero energy, namely to $\delta_l^-(0) = (n_l^- + q_l^-)\pi$, where $n_l^-$ is the number of bound electron states with angular momentum $l$ in the system $e^- + A@C_N$, while $q_l^-$ is the number of bound electron states with the angular momentum $l$ in the target itself [1-3]. Therefore, the behavior of phases as functions of $E$ is qualitatively different in cases when we consider as projectiles $e^+$ and $e^-$ in collisions with $A@C_N$ as well as $C_N$.

We treat the formation of $\tilde{P}s$ in a way, similar to that developed for the case of $e^+ + A$ in [4, 5]. In general, to consider strong $e^+e^-$ interaction in the field of a residual ion demands at least to solve a three-body problem. Instead, much simpler approach was suggested long ago in [4] and developed in e.g. [5]. In this approach the formation of $\tilde{Ps}$ during the collision process is taken into account shifting the total energy of intermedium $e^+e^-$ state by free positronium binding energy $(-I_{Ps})$, where $I_{Ps}$ is $Ps$ ionization potential. Other approximate approaches that take into account the $e^+e^-$ interaction (see e.g. [7]) are much more complex.

We calculate the polarization potential $\hat{V}_A^{pol,+}(r)$ in the random phase approximation with exchange (RPAE) frame [8], while $C_{60}$ represent by a static square well potential $W_F^+(r)$. Its parameters are chosen similar to $W_F^-(r)$, but with an opposite sign. As to $W_F^-(r)$, its parameters are chosen to represent the experimentally known electron affinity of $C_{60}^-$, and low- and medium energy photoionization cross-sections of $C_{60}$ [9]. Along with $W_F^+(r)$ we take into account the polarization potential $V_F^{pol,+}(r)$ of the fullerene. We pay special attention to the development of an approximation that permits to calculate the interference polarization potential $\hat{V}_{FA}^{pol}(r)$, and corresponding phase-shifts as well as cross-sections.

Entirely, the problem of $e^+$, as well as $e^-$ scattering upon endohedrals is very complicated. Our aim is to present the first steps in this direction.

**2.** In order to obtain $e^+$ scattering phases for a spherical endohedral, one has to solve numerically the equations for the radial parts of the one-electron wave functions $P_{El}^{A@C_N,+}(r)$ [2]

$$\left(-\frac{1}{2}\frac{d^2}{dr^2}+\frac{Z}{r}-V_H(r)+W_F^+(r)+V_F^{pol,+}(r)+V_{FA}^{pol,+}(r)\frac{l(l+1)}{2r^2}-E\right)P_{El}^{A@C_N,+}(r)=0. \qquad (1)$$

Here Z is the inner atom nuclear charge and $V_H(r)$ is the Hartree potential of the atom $A$. The asymptotic of $P_{El}^{A@C_N,+}(r)$ determines the scattering phase $\delta_l^{A@C_N,+}(E)$

$$P_{El}^{A@C_N,+}(r)\big|_{r\to\infty}\approx\frac{1}{\sqrt{\pi p}}\sin\left[pr-\frac{\pi l}{2}+\delta_l^{A@C_N,+}(E)\right]. \qquad (2)$$

Here $p^2=2E$.

If one neglects $\left[Z/r-V_H(r)+V_{FA}^{pol,+}(r)\right]$ in (1), the equations (1) and (2) determine scattering function and phase shift of a positron on an empty fullerene.

**3.** More details on how to obtain scattering phases numerically one can find in [10]. We assume that $W_F^+(r)=-W_F^-(r)$. Potential $W_F^-(r)$, as in [1, 2],: is a square well with the depth 0.52 and inner $R_1$ (outer $R_2$) radiuses equal to $R_1=5.26$ ($R_2=8.17$). The potential $V_F^{pol,+}(r)$ is determined by the following expressions, similar to that in [1, 2]:

---

[2] We employ the atomic system of units, with electron mass $m$, electron charge $e$, and Planck constant $\hbar$ equal to 1.

$$V_F^{pol,+}(r) = -\frac{\alpha_F^+}{2(r^2 + b^{+2})^2}\,, \qquad (3)$$

$$\alpha_F^+ \equiv \alpha_F^-(I_{Ps}) = \frac{c}{2\pi^2} \int\limits_{I_F}^{\infty} \frac{\sigma_F(\omega')d\omega'}{\omega'^2 - I_{Ps}^2}\,. \qquad (4)$$

Here $\sigma_F(\omega)$ is fullerene photoionization cross-section. The energy shift $I_{Ps}$ takes into account, as it was suggested in [4, 5] and discussed above, the virtual $\tilde{P}s$ formation. As to the expression for $V_F^{pol,-}(r)$, it includes $\alpha_F^-(0) \equiv \alpha_F$, where $\alpha_F$ is the static dipole polarizability of the fullerene that for $C_{60}$ and a number of other fullerenes is measured and/or calculated. The parameters $b^+$ and $b^-$ are both of the order of fullerenes radius and the results of calculations are not too sensitive to their precise values. This is why we choose $b^+ = b^- = (R_1 + R_2)/2 \equiv R$. The simple expression for $V_F^{pol}(r)$ is widely used in atomic scattering calculations (see e.g. [8])

In principal, the polarization potentials are energy-dependent and non-local. We have an experience to determine it for electron-atoms scattering employing perturbation theory in inter-electron interaction and limiting ourselves by second order perturbation theory in the interaction of incoming and target electrons (see [8] and references therein). Similar approach is valid for positron scattering.

We solve the equation (1) in the integral form and in energy representation, where for partial wave $l$ it looks like (see Chap 3 of [8] and references therein):

$$\left\langle E\ell \,|\, \hat{\bar{\Sigma}}^l\left(E_1\right) \,|\, E'\ell \right\rangle = \left\langle E\ell \,|\, \hat{\Sigma}^l\left(E_1\right) \,|\, E'\ell \right\rangle + \sum_{E''} \left\langle E\ell \,|\, \hat{\Sigma}^l\left(E_1\right) \,|\, E''\ell \right\rangle \frac{1}{E_1 - E'' + i\delta} \left\langle E''\ell \,|\, \hat{\bar{\Sigma}}^l\left(E_1\right) \,|\, E'\ell \right\rangle, \ (5)$$

where the sum over $E''$ includes also integration over continuous spectrum.

The polarization interaction $\hat{\Sigma}(E)$ leads to an additional scattering phase shift $\Delta\delta_l(E)$ that is connected to the diagonal matrix element of (5):

$$e^{i\Delta\delta_l^+(E)} \sin\Delta\delta_l^+(E) = < E\ell \,\|\, \hat{\bar{\Sigma}}_l^+(E) \,\|\, E\ell >. \qquad (6)$$

Instead of semi-empirical potentials, we employ here the many-body theory approach with its diagrammatic technique [8]. The matrix elements $\left\langle E\ell \,|\, \hat{\Sigma}^l\left(E_1\right) \,|\, E'\ell \right\rangle$ have the name "irreducible self-energy part of the single-particle Green's function".

This approach accounts for non-locality and energy dependence of the polarization interaction, but to be accurate enough require inclusion of sufficient number of diagrams' sequences. Note that the phases determined using equations (1) and (2) or (5) and (6) are the same (see e.g. [8]). This was checked by us also pure numerically, by applying both procedures to the case of an empty fullerene that led to identical results.

Just as in the $e^-$ case we limit ourselves with the same diagrams [8], but neglect exchange and add $e^+e^-$ interaction. This leads to (7) where the heavy line stands for $e^+$. In (3) it corresponds to $V_F^{pol,+}(r)$. The shaded oval denotes the $e^+e^-$ interaction.

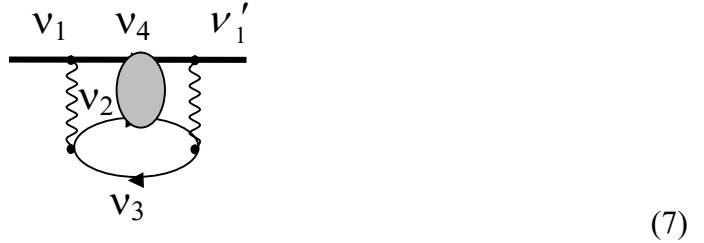

(7)

The wavy line stands for the interelectron interaction. In (7) we use the notations $\nu_i = E_i l_i$. A line, directed to the right (left), denotes electron (vacancy). This diagram automatically includes some infinite series in electron-vacancy interaction (See Chap. 3 in [8, 11]).

**4.** When we consider a positron colliding with an endohedral, one has to take into account the contribution of the interaction between atomic and fullerenes electrons. Diagrams (8) present examples of such interaction that contributes to $V_{FA}^{pol,+}(r)$:

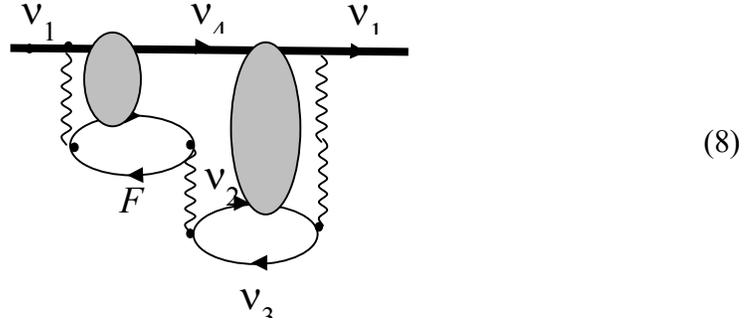

(8)

Here $F$ denotes the fullerenes shell virtual excitations.

Considering the insertion of fullerenes shell virtual excitation and estimating the corresponding contributions, one has to have in mind that between the essential for the scattering process projectile distance $r_p$, the fullerene radius, $R_F$ and the atomic radius, $r_A$, the following inequality exists $r_p > R_C > r_A$. To simplify the problem of taking into account the mutual influence of atomic and fullerenes electron, we enforce this inequality into $r_p \gg R_C \gg r_A$. This permits to limit ourselves by correcting the dipole interelectron interaction only, substituting the Coulomb interelectron potential in the following way $1/|\mathbf{r_1} - \mathbf{r_2}| \to \mathbf{r_1}\mathbf{r_2}/r_2^3$ for $r_1 \ll r_2$. The alteration of the long-range dipole interelectron interaction $V_1$ matrix elements is taken into account similarly to the inclusion of the polarization factor in photoionization of endohedrals as it was demonstrated in [12]. So, we correct them approximately, substituting $|V_1|^2$ by

$$|V_1|^2 \to \left| V_1 \left[ 1 - \alpha_F^+ (E_{\nu_1} - E_{\nu_4}) / R_F^3 \right] \right|^2 \equiv \left| V_1 \left[ 1 - \alpha_F^- (E_{\nu_1} - E_{\nu_4} + I_{Ps}) / R_F^3 \right] \right|^2 \qquad (9)$$

The contribution of the shaded oval or $e^+e^-$ interaction is taken into account by introducing $I_{Ps}$ into the energy argument of the polarizability. We left unchanged other than dipole components of interaction matrix elements.

**5.** In [12] (see also [8]) we have calculated the polarizability $\alpha_F^-(\omega + I_{Ps})$. Details on how to find $\left\langle E\ell \mid \hat{\Sigma}_l^+(E_1) \mid E'\ell \right\rangle$ and to solve equations (3) and (4) one can find in Chap. 3 of [8].

In this Letter we concentrate on presenting qualitative differences between positron and electron low-energy scattering upon endohedral $A@C_{60}$. This is why we pay so much attention to the case a simple atom $He$ stuffed inside $C_{60}$.

In Fig. 1 we present the results of calculations for $s$ scattering phases and cross-sections in $e^+$ and $e^-$ by $He$, $C_{60}$, and $He@C_{60}$. We see that the influence of the inner atom $He$ is relatively small, just as small is the low-energy cross-section in collision $e^+ + He$ as compared to that of $e^- + He$. It is a consequence of the fact that for $He$ the repulsive $W_F^+(r)$ potential almost entirely compensates the effect of attractive $V_F^{pol,+}(r)$ potential. Results for the phase $s$ in $e^- + He$ demonstrate the additivity of phases, according to which the phase upon $He@C_{60}$ is equal to the sum of scattering phases upon $He$ and $C_{60}$ [1].

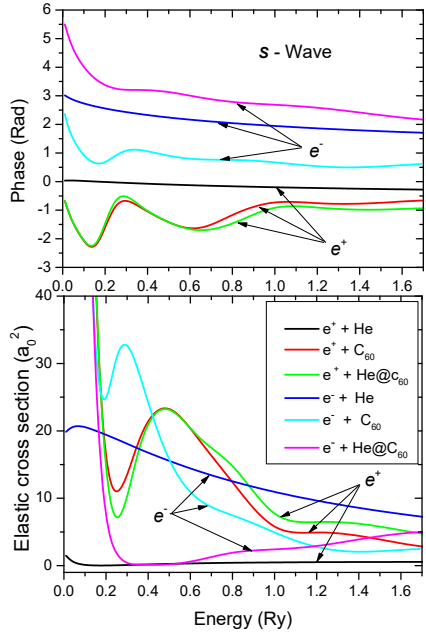

Fig. 1. Role of $e^+e^-$ interaction upon phases and cross-sections of $e^+$ scattering on $He$, $C_{60}$, and $He@C_{60}$, a) $s$-phase with and without $I_{Ps}$ shift, b) Partial $s$ cross-sections of $e^+$ with and without $I_{Ps}$ shift.

The partial $s$-wave cross-section at very low energy is dominated by the $C_{60}$ contribution. The cross-section of $e^+$ upon $He@C_{60}$ has a deep and narrow Ramsauer-type minimum. The structure is much more prominent than in $e^-$ case. In is remarkable that a big maximum at 0.3 Ry in $e^- + C_{60}$ due to quantum effects entirely disappears in $e^- + He@C_{60}$ collision.

Fig. 2 depicts the results for the $p$-phase and $p$

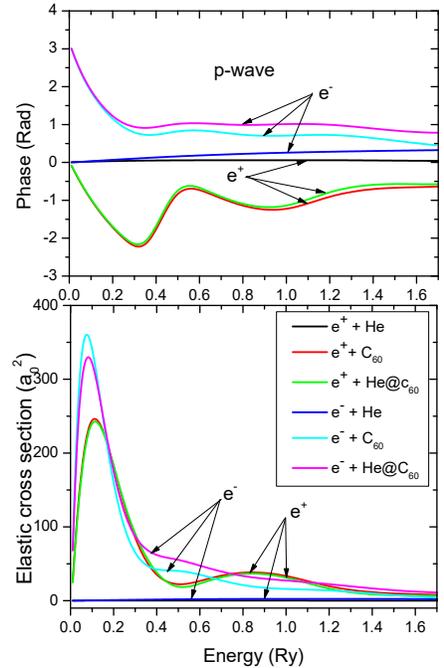

Fig. 2. Role of $e^+e^-$ interaction upon $p$-phase and cross-sections of $e^+$ scattering on $He$, $C_{60}$, and $He@C_{60}$, a) $p$-phase with and without $I_{Ps}$ shift, b) Partial $s$ cross-sections of $e^+$ with and without $I_{Ps}$ shift.

partial cross section. For $e^+$ case the phase is, naturally, much smaller than for $e^-$ and the role of *He* in $He@C_{60}$ is also small. In the $e^-$ case the scattering phase on *He* is big and the additivity rule for phases is valid.

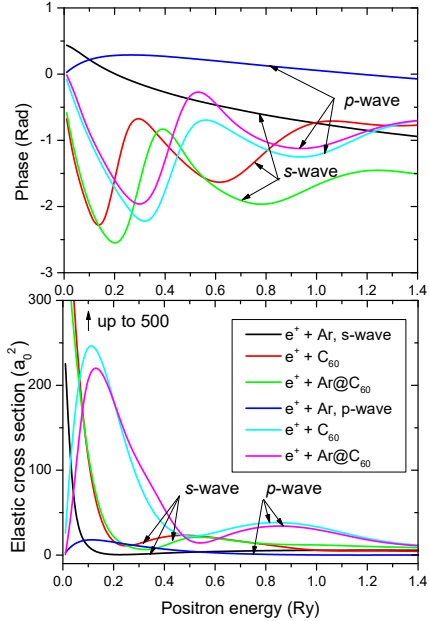

Fig. 3. Comparison of phases and cross-sections of $e^+$ scattering upon *Ar*, $C_{60}$, and $Ar@C_{60}$ a) s and p scattering phases, b) Partial $s$ and $p$ scattering cross sections.

The $p$-partial cross-section for both $e^+$ and $e^-$ has a strong maximum at about 0.1 Ry, the absolute value for $e^-$ being 1.5 times bigger than for $e^+$. The $e^+$ cross-section has a Ramsauer-type deep minimum located a little bit higher for $p$-wave, at 0.5 Ry, than for $s$-wave. For $e^-$ such a minimum is absent.

Fig 3 present the phases and partial cross sections of $e^+$ and $e^-$ low-energy elastic scattering upon *Ar*, $Ar@C_{60}$. For convenience of the reader, Fig. 3 includes the curves that represent the collision of $e^+$ with $C_{60}$. It appeared that while for the $p$-wave the rule of additivity is valid, it does not work for the $s$-wave.

The absence of exchange with the target constituents and lack of the real bound states of $e^+$ with considered targets dictates that the all scattering phases for the zero projectile energy has to be equal to zero. The fact that the s-phase has in Ar a non-zero value at low energy signals the presence of a scattering resonance close to zero energy. And indeed, the $s$-cross section has a powerful maximum (up to 500), and then, at about 0.3 Ry, has a very deep Ramsauer minimum. As for the $p$-wave, its maximum is at about 0.15Ry and is two times lower than for s-wave. The role of *Ar* inside $C_{60}$ is noticeable for both considered partial waves – $s$ and $p$.

**6.** In this Letter we present an approach that permits to study the positron scattering upon endohedrals $A@C_{60}$. This approach permits to take into account the polarization of $A$ and $C_{60}$ as well as the interference of their polarizabilities. Our approach also accounts for strong positron – target electron interaction that leads to the concept of virtual positronium formation in the scattering process.

For two objects, namely $He@C_{60}$ and $Ar@C_{60}$, we performed numeric calculations that demonstrate a prominent role of the atom $A$ in the formation of the scattering cross-section and a big difference between the positron and electron scattering. As in the case of electron scattering upon endohedral, we observe a big difference for the cross-sections of fullerene and endohedral. It means that the cross section is determined by quantum mechanics, losing its most prominent feature – dependence at low scattering energy only upon the size of the target.

The observed resonance structure of the cross-sections deserves experimental investigation, since it permits to shed light on the difference in potentials, by which an endohedral acts upon a projectile – positron or electron.

We do believe that the suggested method and presented results of concrete calculations will stimulate theoretical and experimental research of low-energy elastic scattering of positrons by endohedral atoms.

# **Figures** (color on line)

## **Figure captions**

Fig. 1. Role of $e^+e^-$ interaction upon $s$ - phase and cross-sections of $e^+$ scattering on *He*, $C_{60}$, and *He*@$C_{60}$, a) $s$-phase with and without $I_{Ps}$ shift, b) Partial $s$ cross-sections of $e^+$ with and without $I_{Ps}$ shift.

Fig. 2. Role of $e^+e^-$ interaction upon $p$ - phase and cross-sections of $e^+$ scattering on *He*, $C_{60}$, and *He*@$C_{60}$, a) $p$-phase with and without $I_{Ps}$ shift, b) Partial $s$ cross-sections of $e^+$ with and without $I_{Ps}$ shift.

Fig. 3. Comparison of phases and cross-sections of $e^+$ scattering upon *Ar*, $C_{60}$, and *Ar*@$C_{60}$ a) s and p scattering phases, b) Partial $s$ and $p$ scattering cross sections.

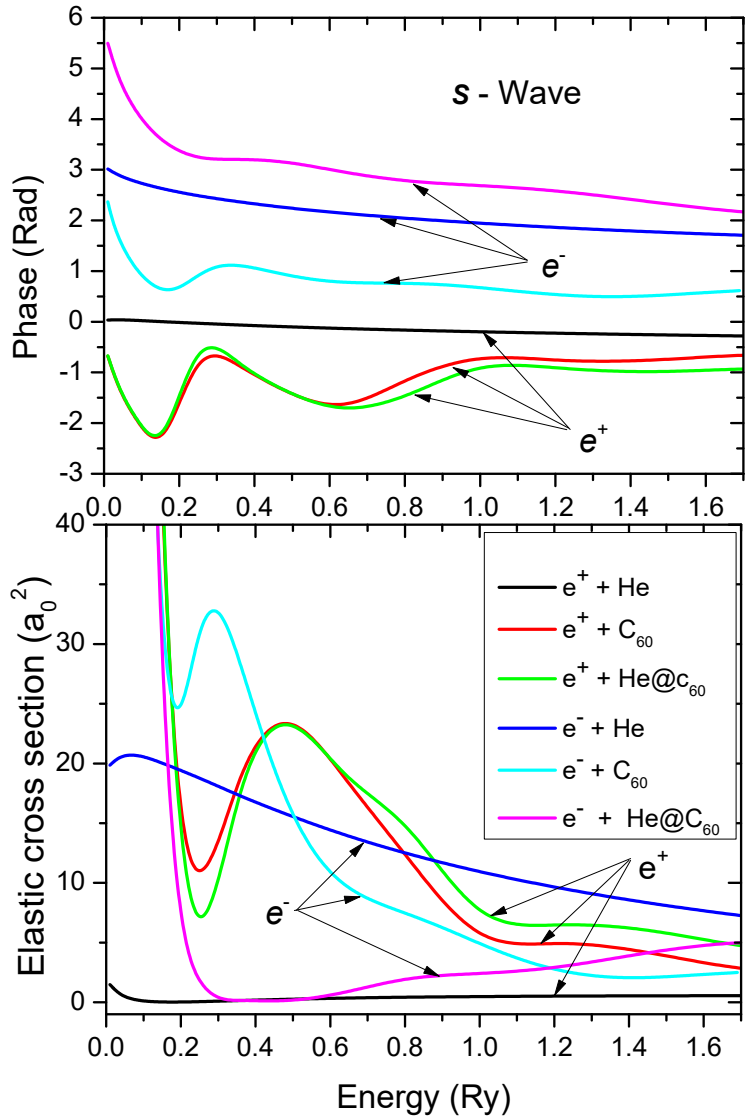

Fig. 1. Role of $e^+e^-$ interaction upon phases and cross-sections of $e^+$ scattering on $He$, $C_{60}$, and $He@C_{60}$, a) $s$-phase with and without $I_{Ps}$ shift, b) Partial $s$ cross-sections of $e^+$ with and without $I_{Ps}$ shift.

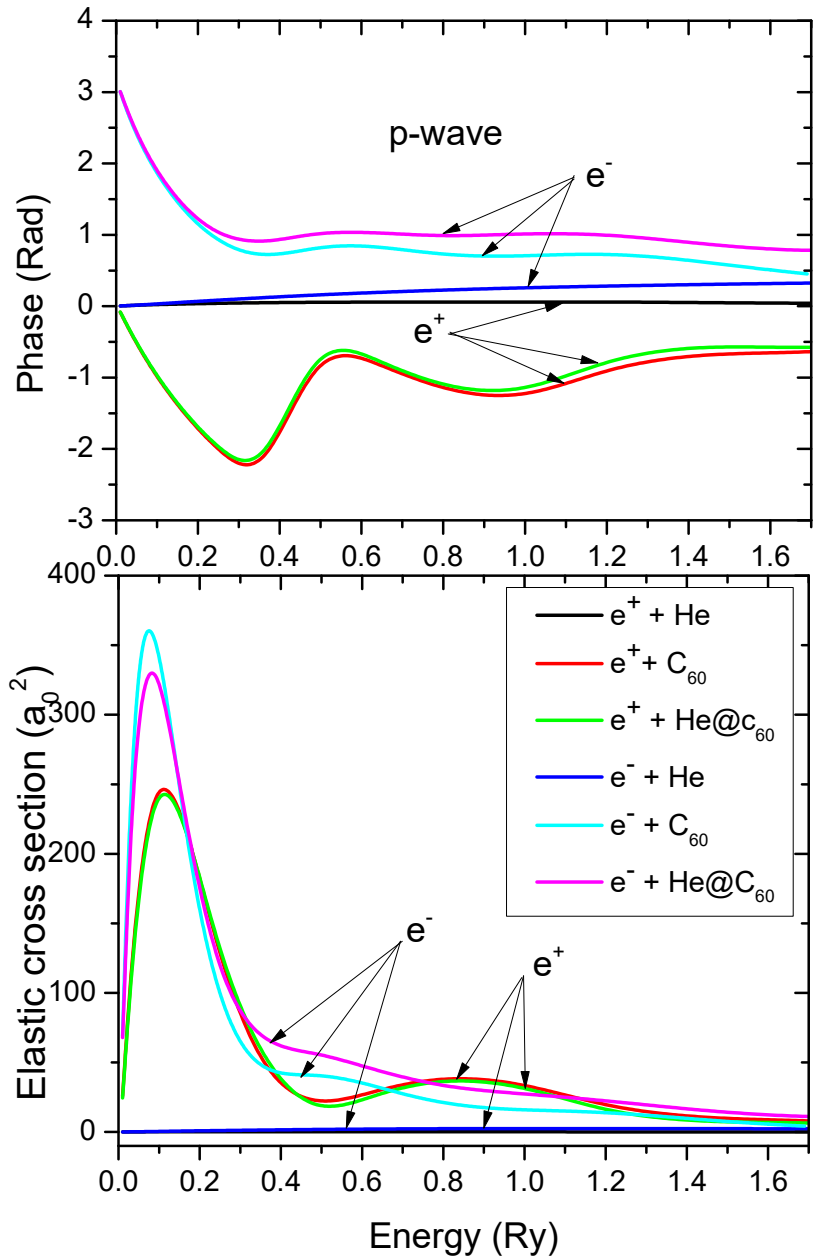

Fig. 2. Role of $e^+e^-$ interaction upon $p$ - phase and cross-sections of $e^+$ scattering on $He$, $C_{60}$, and $He@C_{60}$, a) $p$-phase with and without $I_{Ps}$ shift, b) Partial $s$ cross-sections of $e^+$ with and without $I_{Ps}$ shift.

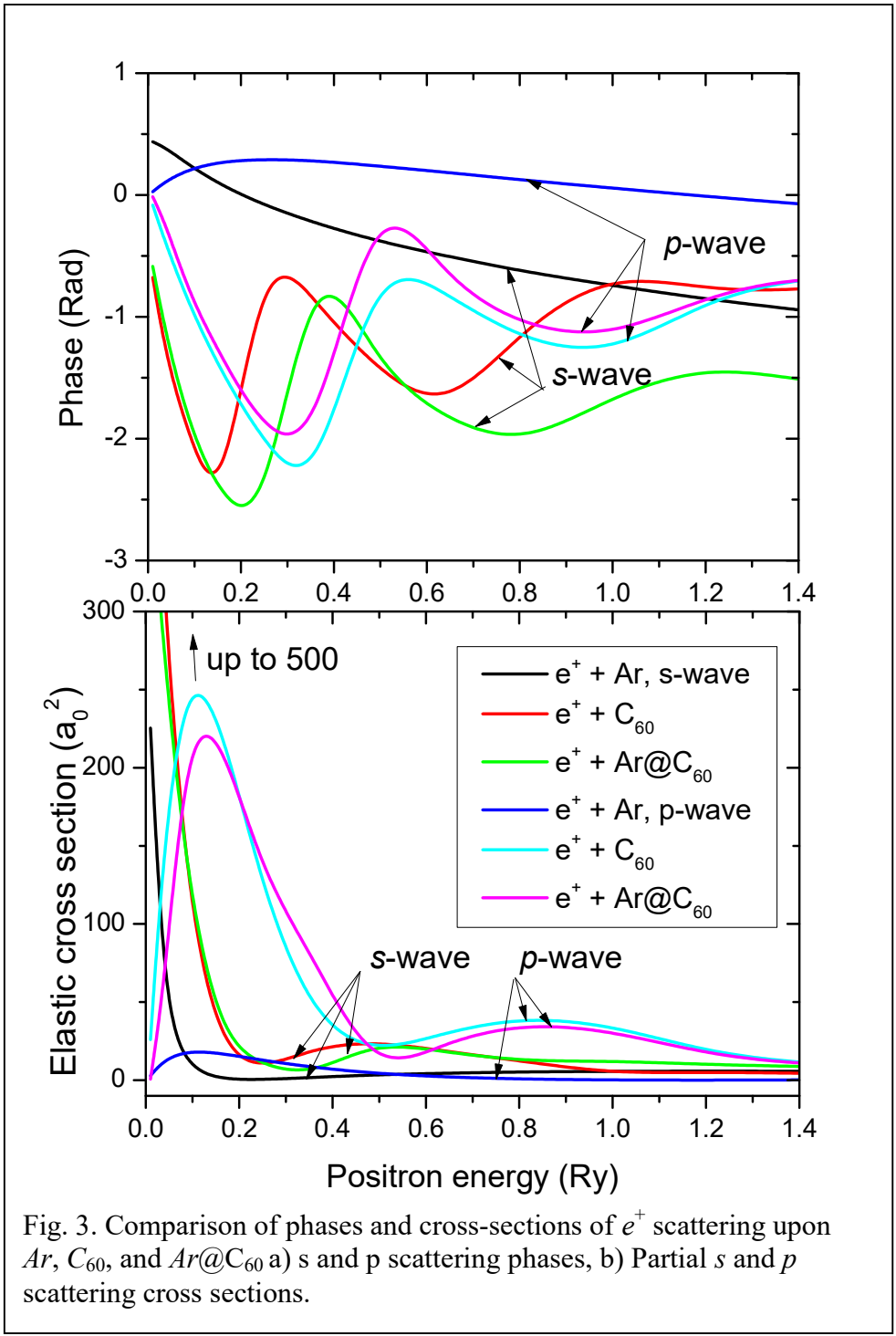

Fig. 3. Comparison of phases and cross-sections of $e^+$ scattering upon *Ar*, $C_{60}$, and *Ar*@$C_{60}$ a) s and p scattering phases, b) Partial *s* and *p* scattering cross sections.